\documentclass{iaus}
\usepackage{graphicx,amssymb,natbib}

\title[IAUS266.~~NGC\,604: near-infrared Gemini/NIRI imaging] 
{A deep dive into NGC\,604 with Gemini/NIRI imaging}

\author[C. Fari\~na et al.] 
{Cecilia Fari\~na$^{1, 2}$, Guillermo L. Bosch$^{1, 2}$
 \and Rodolfo H. Barb\'a$^{3, 4}$}

\affiliation{$^1$Facultad de Ciencias Astron\'omicas y Geof\'{\i}sicas - Universidad Nacional de La Plata, \\ Paseo de Bosque S/N, 1900 La Plata, Argentina \\ email: {\tt ceciliaf@fcaglp.unlp.edu.ar} \\[\affilskip]
$^2$ IALP - CONICET, Argentina \\ [\affilskip]
$^3$ ICATE - CONICET, Argentina \\  [\affilskip]
$^4$ Departamento de F\'{\i}sica, Universidad de La Serena, Benavente 980, La Serena, Chile}

\pubyear{2010}
\volume{266}  
\pagerange{}
\jname{Star clusters: basic galactic buildng blocks}
\editors{R.de Grijs \& J. R. D. L\'epine}
\begin{document}

\maketitle

\begin{abstract}
The giant {\sc Hii} region NGC\,604 constitutes a complex and rich population to studying detail many aspects of massive star formation, such as their 
environments and physical conditions, the evolutionary 
processes involved, the initial mass function for massive stars and 
star-formation rates, among many others. Here, we present our first results of a near-infrared study of NGC\,604 performed with NIRI images obtained with {\it Gemini North}. Based on deep {\it J\,H\,K} photometry, 164 sources showing infrared excess were detected, pointing to the places where we should look for star-formation processes currently taking place. In addition, the color-color diagram reveals a great number of objects that could be giant/supergiant stars or unresolved, small, tight clusters. A extinction map obtained based on narrow-band images is also shown.\\

\end{abstract}

\firstsection 
\section{Introduction}
NGC\,604 is a giant {\sc Hii} region (GHR) located in an outer spiral arm of M33, at 
a distance of 840 kpc. 
It is the second most luminous {\sc Hii} region in the Local Group, after 30 Doradus 
in the LMC. Both are nearby examples of giant star-forming regions whose individual objects can be spatially resolve for further study. NGC\,604 has been the target of many studies during the past few decades. A brief summary of known facts about NGC\,604 includes the following. This GHR is ionized by a massive, young cluster, with at least 200 O stars (some as early as O3-O4). The cluster does not exhibit a central core distribution. Instead, the stars are widely spread over its projected area in a structure called 'scaled OB association' (SOBA; \cite{Hunter_1996}; \cite{Maiz_2004}; \cite{Bruhweiler_2003}). Wolf-Rayet stars, a confirmed and many candidate red supergiant stars, a luminous blue variable and a supernova remnant are all part of NGC\,604's stellar population (\cite{Conti_1981}; \cite{Dodorico_1981}; \cite{Drissen_1993}; \cite{Diaz_1996}; \cite{Churchwell_1999}; \cite{Terlevich_1996}; \cite{Barba_2009}). The age of the central ionizing cluster has  been determined by different authors as between 3 and 5 Myr (\cite{Gonzalez_2000}; \cite{Bruhweiler_2003}; \cite{Diaz_1996}; \cite{Hunter_1996}; \cite{Relano_2009}).\\
 The interestellar medium reveals a complex structure with a high-excitation 
central region (made up of multiple two-dimensional structures), asymmetrically 
surrounded by a low-excitation halo. The whole region shows a very complex 
geometry of cavities, expanding shells and filaments, as well as dense molecular regions. All of these structures show different kinematic behaviour (\cite{Maiz_2004}; \cite{Tenorio-Tagle_2000}; \cite{Sabalisck_1995}; \cite{Relano_2009}). \\ 
Aiming to characterize the youngest stellar population and its environment, we 
performed near-infrared (NIR) photometry ({\it J\,H\,K}) and analysed narrow-band images in 
Pa$\beta$, Br$\gamma$ and H$_2$(2-1). Taking into account that NIR 
observations are less affected by dust extinction characteristic in star-fomation environments, we can take a deep dive into NGC\,604 to study those 
very young objects which are still immersed in their parental clouds at the sites of current star formation.

\section{Images and data processing}

The images were obtained with the Near InfraRed Imager and Spectrometer (NIRI) 
at {\it Gemini North}. The resulting plate scale is 0.117 arcsec\,pixel$^{-1}$, 
with a field of view of 120 $\times$ 120 arcsec$^2$. The filters used and their main 
characteristics are listed in Table 1.

\begin{table}
\caption{Main characteristics of broad-band and narrow-band filters used in our Gemini/NIRI observations.}
\label{tab1}
 \begin{tabular}{ccc|ccc} 
 \multicolumn{3}{c|}{Broad bands} & \multicolumn{3}{c}{Narrow bands}\\   
Filter&Central $\lambda$ ($\mu$m)& Coverage($\mu$m)& Filter&Central $\lambda$ ($\mu$m)& Coverage($\mu$m)\\ \hline
J & 1.25 & 0.97-1.07 & Pa$\beta$ &1.282 &$\sim$ 0.1 \\
H & 1.65 & 1.49-1.78 & Br$\gamma$& 2.16 &$\sim$ 0.1\\
K$_s$ & 2.15 & 1.99-2.30 & H$_2$(2-1) &2.24 & $\sim$ 0.1\\ 
\end{tabular}
\end{table}

The images were taken under excellent seeing conditions, on average $\sim 0.35"$ in the J, H, and K$_s$ images. A set of approximately 10 individual 
exposures was taken in each band to combine into the final image. Data reduction and processing were performed with specific tasks using the {\sc gemini-niri iraf} package. Images were sky subtracted and flat fielded and short darks exposures were used to identify bad pixels. Stellar magnitudes were obtained by point-spread-function (PSF) fitting in crowded fields using {\sc daophot} software (Stetson 1987) in {\sc iraf}. Although a standard procedure, PSF construction and fitting involves an iterative and careful 
process in which several tries were made to get the best results. The 
effective area covered by our photometry is $\sim$ 107 $\times$ 107 arcsec$^2$ ($\sim$ 430 $\times$ 430 pc$^2$ at the distance of M33). The average photometric errors are 0.09, 0.11 and 0.21 mags in the H, J and K$_s$ filters, respectivelly, and the completeness limits are 22 mags in J and 21 mags in H and K$_s$. Magnitudes in the individual filters were matched in a unique list containing 5566 objects in the field in which all three J, H and K$_s$ magnitudes were measured. Astrometry was derived using 35 objects in common in our field of NGC\,604 and the GSC-II Catalog, Version 2.3.2 (2006) in the ICRS, Equinox J2000.0.\\

\section{Results and discussion}

\begin{figure}[]
\begin{center}
 \includegraphics[width=0.49\textwidth]{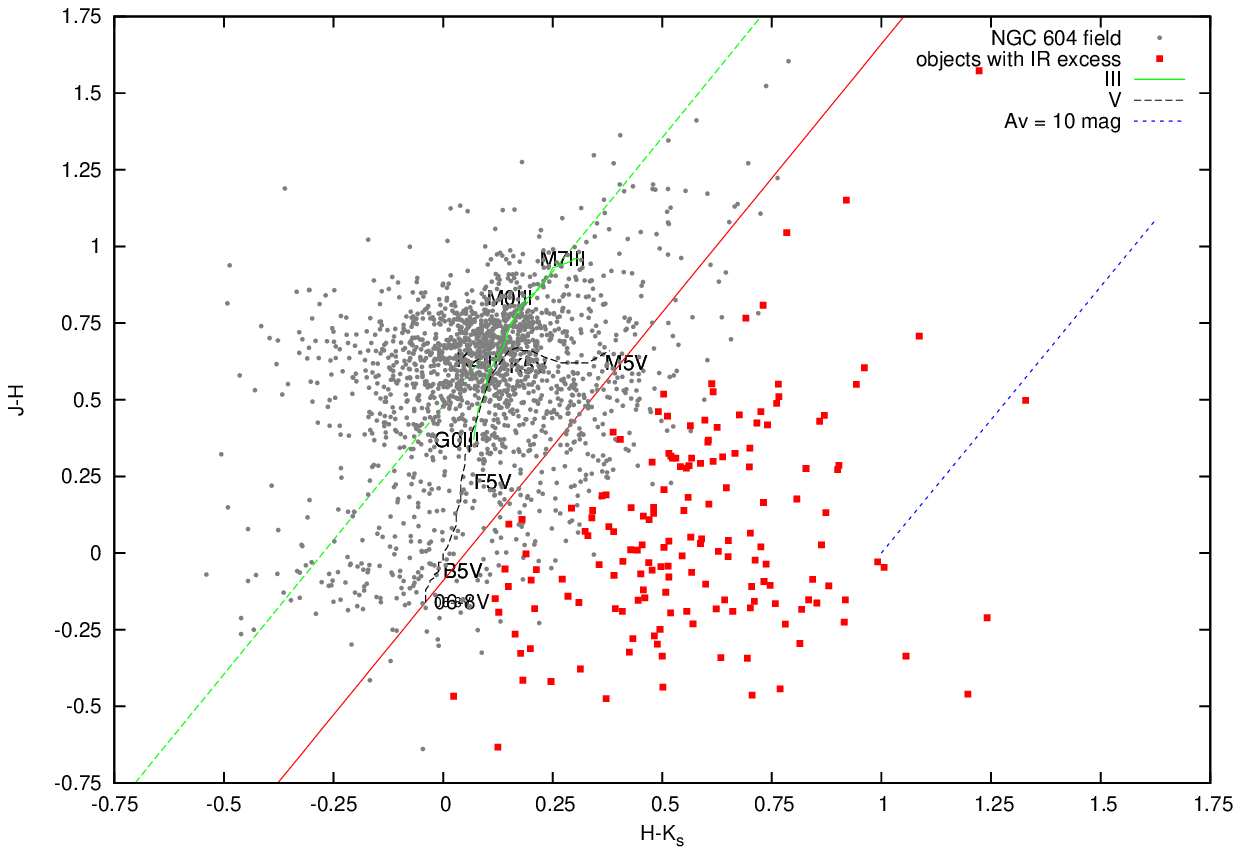} 
 \includegraphics[width=0.49\textwidth]{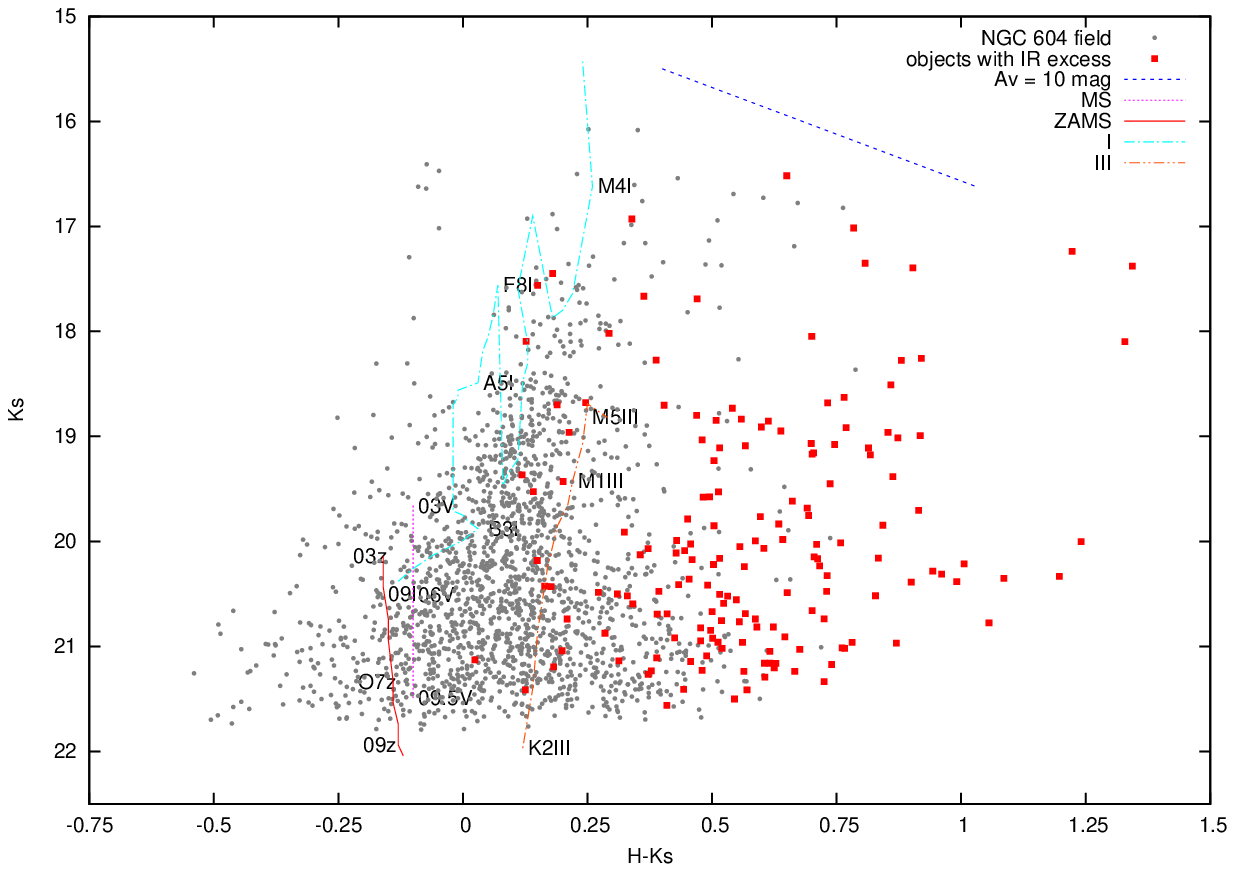}
 \caption{Color-color {\it (left)} and color-magnitude {\it (right)} diagrams for objects observed in a circular area centered on NGC\,604. Red squares are objects that show an IR excess.}
   \label{fig1}
\end{center}
\end{figure}

\begin{figure}[b]
\begin{center}
 \includegraphics[width=0.45\textwidth]{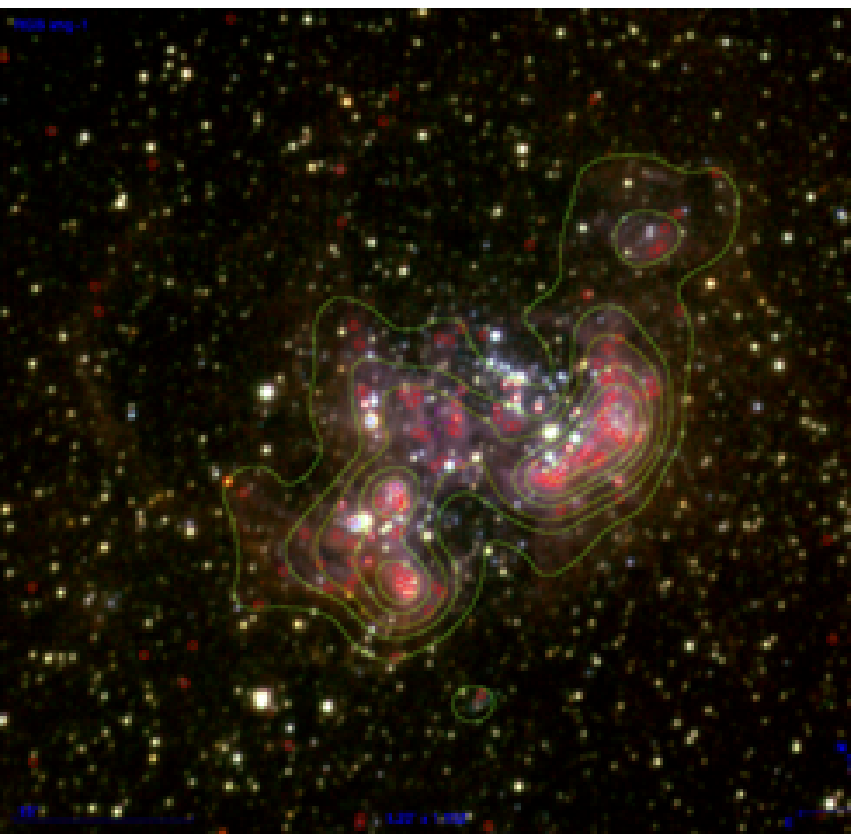} 
 \includegraphics[width=0.53\textwidth]{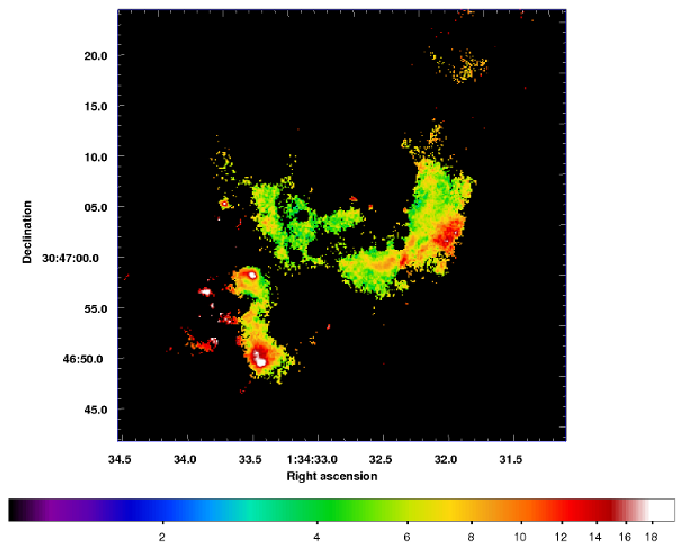}
 \caption{{\it (left)} Three-color (J, H and K$_s$) composite image of NGC\,604 with 8.4 GHz radio-continuun countours overlaid (adapted from \cite{Churchwell_1999}). Red circles are objects that show IR excess in our photometry. {\it (right)} Extinction map  from Br$\gamma$/H$\alpha$, regions with higher extinction are in red/white.}
   \label{fig2}
\end{center}
\end{figure}

The resulting color-magnitude (CM) and color-color (CC) diagrams are shown in Figures 1 and 2, respectively. We have included $\sim$ 2000 selected objects, located within a radius of 48 arcsec ($\sim$ 200 pc) and centered on  NGC\,604, meeting a certain photometric quality level (magnitude error $\leq$ median(error)+ 1.0 $\times$ standard deviation(error)). Red symbols in both plots are objects that lie on the right side of the reddening line for a O6-O8 V star.For each of these objects, the error in (H-K$_s$) color is smaller than its distance to the reddening line, so that we can ensure that they undoubtedly show an IR excess.\\ 
Among the objects with intrinsic IR excesses in GHRs we expect to find Wolf-Rayet 
stars, early supergiants and Of stars, and massive young stellar objects (MYSOs). Many of these types of objects (or candidates) have already been found in NGC\,604. Based on our deep {\it J\,H\,K} photometry we found a total of 164 objects that show IR excesses. As can be expected, $\sim$ 70\% of these lie in a small area near the region's center and a large proportion are tightly grouped in regions coincident with the radio-continuum peaks at 8.4 GHz from \cite{Churchwell_1999}, as shown in Figure 3 (left panel). These results are in complete agreement with the analysis of \cite{Barba_2009} based on HST/NICMOS NIR images, where MYSO candidates also appeared aligned with the radio peak structures.\\
What we found supports the idea mentioned by many authors, that those areas may be embedded star-forming regions, also taking into account that regions coincident with the radio knots show conditions of high temperature and density (\cite{Tosaki_2007}), and in two of them \cite{Maiz_2004} identified compact {\sc Hii} regions. Most objects with IR excesses will be the targets for further observing programs (making use of the integral-field spectroscopic facilities at {\it Gemini North}) to elucidate their nature and derive accurate properties by means of spectroscopic study.
Also, those objects that are located at the bright, massive end on the CM diagram
deserve further observation and study.\\
On the basis of our narrow-band images we generated an extinction map, derived from the observed variations of the Br$\gamma$ to H$\alpha$ ratio (with the H$\alpha$ image taken from \cite{Bosch_2002}). The color scale used in the map shows regions with higher extinction in red/white. The present and future results of our NIR study will be placed  in the context of previous studies of NGC\,604 to complete the picture of the overall formation and evolution scenario of this GHR.\\

{\bf Acknowledgements}\\
RB acknowledges partial support from the Universidad de La Serena, project DIULS CD08102.

\end{document}